\documentclass[aps,pre,reprint,10pt]{revtex4-2}

\usepackage{graphicx,amsmath,amssymb,dcolumn,physics,color,bm}

\newcommand{\x}{\bm{x}}

\newcommand{\kB}{k_\mathrm{B}}
\newcommand{\mobility}{\bm{\mu}}

\newcommand{\A}{\bm{A}}
\newcommand{\cov}{\bm{C}}
\newcommand{\mean}{\bm{m}}

\begin{document}

\title{Will a Large Complex System be a Maxwell Demon?}

\author{Matthew P. Leighton}
\email{matthew.leighton@yale.edu}
\affiliation{Department of Physics and Quantitative Biology Institute, Yale University, New Haven, CT 06511}
\date{\today}

\begin{abstract}
Emerging evidence suggests that physical systems operating as Maxwell demons, in which some subsystem of a larger system extracts heat energy from its environment in an apparent local violation of the second law, are commonplace throughout biology. Should these findings surprise us, or is Maxwell demon behavior inevitable in sufficiently large complex systems? In this Letter we pose the question of how likely it is that a random stochastic system with many degrees of freedom will operate as a Maxwell demon, considering null models for both continuous and discrete random dynamics. Our results show the probability of a finding a demon decreases at least exponentially, and in some cases double-exponentially, with the number of degrees of freedom, ultimately suggesting that large complex demons can only arise through a process of selection.
\end{abstract}

\maketitle

The specter of Maxwell’s demon lies all around us~\cite{binder2011life,bejan2014maxwell}, with sightings reported across diverse fields from economics~\cite{morck1995value,hanappi2003maxwell}, to evolutionary biology~\cite{adami2000evolution,krakauer2011darwinian}, to cell sensing and signaling~\cite{tu2008nonequilibrium,ito2015maxwell,boel2019omnipresent}, to molecular biophysics~\cite{flatt2023abc,leighton2024information,jiménez2025information,buisson2025hunting,leighton2024flow}. The core phenomenon of a Maxwell demon is that some subsystem of a larger thermodynamic system appears to locally violate the second law by taking in heat energy from its environment~\cite{Maxwell1867,Leff2003_Maxwells,Parrondo2015_Thermodynamics}. This apparent violation is rectified by a flow of information between subsystems, whereby the subsystem taking in heat energy must in turn decrease its shared information with the rest of the system, since this information is itself a thermodynamic resource~\cite{horowitz2014thermodynamics,horowitz2015multipartite,Parrondo2015_Thermodynamics,leighton2024flow}.

The finding of Maxwell demon behavior, especially in biological settings, has been used to argue that such systems must have evolved to use information in order to leverage fluctuations in their environment~\cite{buisson2025hunting,jiménez2025information}. But can we really draw such conclusions about evolved behavior from the mere observation of a Maxwell demon? Biological systems are complex, composed of many interacting degrees of freedom. Perhaps in a sufficiently large complex system, it is inevitable that at least one degree of freedom will appear to locally violate the second law. It thus behooves us to ask the question: is the presence of Maxwell demon behavior surprising, or is it simply something we should generically expect to find in sufficiently large complex systems?

Inspired by the seminal work of Robert May~\cite{may1972will}, we ask whether a random complex system is likely to behave as a Maxwell demon. To address this question, we formulate null models of random multipartite systems obeying both continuous Langevin dynamics and discrete master equation dynamics, spanning both near-equilibrium and far-from-equilibrium regimes. We show both analytically and numerically that the probability of such random systems operating as Maxwell demons decreases at least exponentially, if not faster, with the number $N$ of degrees of freedom. Our findings suggest that observing a Maxwell demon in the wild really should come as quite the surprise.

\emph{Stochastic thermodynamics of multipartite systems}.---As a paradigm, consider multipartite stochastic systems with $N$ degrees of freedom, denoted $\x=\{x_k\}_{k=1}^N$. We assume that each subsystem, characterized by a single degree of freedom, is in contact with a thermal reservoir at temperature $T$. Here multipartite means that each subsystem can independently exchange heat energy with the environment; for discrete systems this means that transitions between states can only change one degree of freedom at a time~\cite{Ehrich2023_Energy}, while for continuous systems this means the diffusion tensor is diagonal~\cite{horowitz2015multipartite}.

To describe the thermodynamics, we focus our attention on nonequilibrium steady states. The subsystems can each exchange work with external work reservoirs at rates $\dot{W}_k$, increase or decrease the system's internal energy at rates $\dot{E}_k$, and exchange heat energy with their thermal reservoir at rate $\dot{Q}_k$. We take the convention that work and heat into the system are positive. These flows of energy are related by both a global first law
\begin{equation}
\sum_{k=1}^N \dot{E}_k = \sum_{k=1}^N \left(\dot{W}_k + \dot{Q}_k\right),
\end{equation}
as well as local first laws that describe energy balance at the level of each individual subsystem~\cite{leighton2025stochastic}:
\begin{equation}
\dot{E}_k = \dot{W}_k + \dot{Q}_k.
\end{equation}

At steady state, the second law of thermodynamics requires that the entropy production rate $\dot{\Sigma}$, equal to the total rate of heat dissipated to external reservoirs by all subsystems, is nonnegative:
\begin{equation}
\dot{\Sigma} = -\sum_{k=1}^N\frac{\dot{Q}_k}{T}\geq 0.
\end{equation}
Individual subsystems must additionally satisfy local second laws, where each subsystem entropy production $\dot{\Sigma}_k$ is given by a sum of heat and information flows~\cite{horowitz2015multipartite}:
\begin{equation}
\dot{\Sigma}_k = -\frac{\dot{Q}_k}{T} - \dot{I}_k\geq 0.
\end{equation}
Here the information flow $\dot{I}_k$ is the rate at which the dynamics of the $k$'th subsystem change the total correlation~\cite{crooks2017measures} (a multivariate generalization of the mutual information) between all degrees of freedom. At steady state the total correlation is constant, so that the information rates of all subsystems must sum to zero,
\begin{equation}
\sum_{k=1}^N\dot{I}_k=0.
\end{equation}

Thus the subsystem laws of thermodynamics permit any given subsystem to take in heat from the environment ($\dot{Q}_k\geq0$) so long as it is accompanied by a sufficiently large information flow $-\dot{I}_k\geq \dot{Q}_k/T$, which in turn comes at the cost of heat dissipation ($\dot{Q}_j<0$) from other subsystems. Without accounting for this information flow, the $k$'th subsystem appears to locally violate the second law, since if observed in isolation it would apparently spontaneously absorb heat from its environment. Here we define a Maxwell demon to be any multipartite system which features at least one subsystem with a positive heat flow, so that there exists a $k\in[1,N]$ such that $\dot{Q}_k>0$. 

This is a relatively weak definition compared to others that have been proposed in the literature. Stronger definitions require, for example, that no energy be exchanged between the subsystems either at the ensemble-averaged level~\cite{sanchez2019nonequilibrium,ciliberto2020autonomous,saha2021maximizing} ($\dot{E}_k=0$) or at the level of individual trajectories~\cite{freitas2021characterizing}.  Nonetheless, our definition includes all demons captured by stricter definitions such as these, and thus probabilities we derive for Maxwell demon behavior will more generally be upper bounds on probabilities under stricter definitions.

Our general question of how likely random complex systems are to be Maxwell demons can now be sharpened into the following: for a random multipartite system with $N$ degrees of freedom, what is the probability that at least one subsystem will feature a positive heat flow from the environment? Answering this question requires null models for random systems, which we now develop for both continuous and discrete dynamics.

\emph{Linear Langevin dynamics}.---As a first null model for random stochastic dynamics, consider a set of $N$ overdamped linear Langevin equations:
\begin{equation}
\dot{\x} = \mobility\left[ \bm{f} - \A\x\right] + \bm{\eta}(t).
\end{equation}
Here $\mobility$ is the mobility tensor (assumed to be diagonal), $\bm{f}$ is a constant force vector, $\A$ is a matrix characterizing linear forces, and $\bm{\eta}(t)$ is zero-mean Gaussian white noise with correlations described by
\begin{equation}
\left\langle \bm{\eta}(t)\bm{\eta}^\top(t')\right\rangle = 2 \mobility\kB T \, \delta(t-t'),
\end{equation}
for Boltzmann's constant $\kB$ and Dirac-delta function $\delta$.

The probability distribution that solves these stochastic differential equations is a multivariate Gaussian distribution with covariance matrix $\cov = \langle \x\x^\top - \langle\mean\rangle\langle\mean\rangle^\top\rangle$ satisfying the differential equation~\cite{ehrich2020coupled}
\begin{equation}
\dot{\cov} = -\mobility\A\cov -  \cov\mobility^\top\A^\top + 2\mobility T. \label{eq:covode}
\end{equation}
So long as the eigenvalues of the matrix $\A$ are all positive, these dynamics will have a unique steady state where $\langle\dot{\x}\rangle=0=\dot{\cov}$. In this case the heat flow into the $k$'th subsystem is~\cite{horowitz2015multipartite}
\begin{equation}\label{eq:continuousheatflow}
\dot{Q}_{k} = \mu_{kk}\left[A_{kk} T - (\bm{A}\bm{C}\bm{A}^\top)_{kk}\right].
\end{equation}

To make progress, we switch to dimensionless units where $T=1$ and $\mobility = \bm{I}$ (assuming equal mobilities), where $\bm{I}$ is the identity matrix, and $\mathrm{Tr}(\A)/N = 1$. We then assume the force matrix can be decomposed as $\A = \bm{I} + \epsilon \bm{M}$, thus assuming equal diagonal terms. Here $\bm{M}$ contains random off-diagonal interaction coefficients (with $M_{kk}=0$), which we take to be independent and normally distributed with mean $0$ and variance $1$, and $\epsilon \ll 1$ is a perturbative parameter. For a bipartite system ($N=2$) only at most one heat flow can be positive, and thus we can solve for the probability that one or the other heat flow is positive analytically, obtaining $p(\mathrm{demon})=1/2$ in the small-$\epsilon$ limit (Appendix A).

For larger $N$, we expand the solution for the covariance matrix to second order in $\epsilon$, obtaining for the heat flow
\begin{equation}
\dot{Q}_k = \frac{\epsilon^2}{2} \sum_{j \neq k} \left( M_{kj}M_{jk} - M_{kj}^2 \right) + \mathcal{O}\left(\epsilon^3\right).
\end{equation}
Defining $\bm{U} \equiv \{M_{kj}\}_{j\neq k}$ and $\bm{V}\equiv \{M_{jk}\}_{j\neq k}$, which are both independent and isotropic random vectors in $\mathbb{R}^{N-1}$, a positive heat flow $\dot{Q}_k>0$ requires $\bm{U}\cdot\bm{V}> |\bm{U}|^2$. Satisfying this condition requires a high degree of alignment between two random vectors, an occurrence which becomes increasingly unlikely in higher dimensions. We exactly calculate the probability of satisfying this inequality in Appendix B, finding that the marginal probability that the $k$'th subsystem features a positive heat flow is
\begin{equation}
p_1 \equiv p\left(\dot{Q}_k>0\right) = \frac{1}{2}I_{\frac{1}{2}}\left(\frac{N-1}{2},\frac{1}{2}\right),
\end{equation}
where $I_x(a,b)$ is the regularized incomplete beta function. For large $N$ this scales as $p_1\sim 2^{-N/2}/\sqrt{N}$.

Using the marginal probability, we can then place upper and lower bounds on the probability that at least one subsystem features a positive heat flow. As an upper bound, consider the case where the events of different subsystems with $\dot{Q}_k>0$ are entirely disjoint. In this case we would see
\begin{equation}
p_\mathrm{UB}(\mathrm{demon}|N) = N p_1,
\end{equation}
which for large $N$ scales as $\sqrt N 2^{-N/2}$. Conversely, if the same events are maximally positively correlated, we would obtain 
\begin{equation}
p_\mathrm{LB}(\mathrm{demon}|N) = p_1.
\end{equation}
Since the second law rules out all $\dot{Q}_k>0$ simultaneously this must be a lower bound, scaling as $2^{-N/2}/\sqrt{N}$. Finally, as an intermediate possibility, the events could be independent (though subject to the global second law requiring at least one heat flow be negative), so that
\begin{equation}\label{eq:independentapprox}
p_\mathrm{independent}(\mathrm{demon}|N) = 1 - (1-p_1)^N + p_1^N,
\end{equation}
scaling as $\sqrt N 2^{-N/2}$, identical to the upper bound.

\begin{figure}[t]
    \centering
    \includegraphics[width=\linewidth]{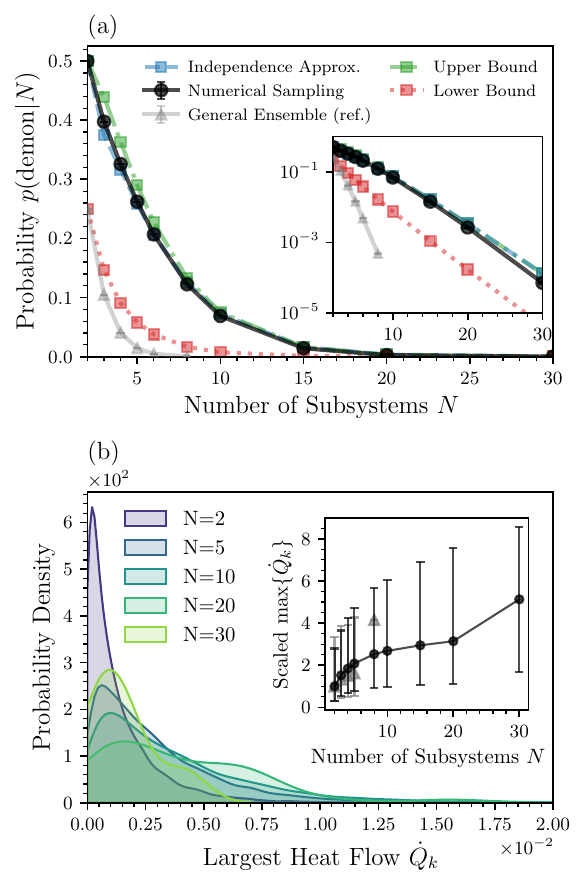}
    \caption{a) Probability that a random stochastic system obeying linear overdamped Langevin dynamics operates as a Maxwell demon. Theoretical lower (red) and upper (green) bounds, as well as the independence approximation (blue). Black points show numerical results from sampling random $\A$ matrices subject to the constraint of all positive eigenvalues (uncertainties are smaller than the displayed points), with $\epsilon=0.1$, while grey points show a more general ensemble (see Appendix E for details). Inset: logarithmic y-axis. b) Distribution of largest observed heat flows in random systems that are Maxwell demons for different numbers $N$ of degrees of freedom. Inset: mean and left/right standard deviations as functions of $N$.
    }
    \label{fig:continuousscaling}
\end{figure}

Numerically sampling from randomly generated interaction matrices $\A$ satisfying the above assumptions as well as the requirement that all eigenvalues be positive, we find a probability of Maxwell demon behavior that falls between the upper and lower bounds derived above, seemingly scaling as $\sqrt N 2^{-N/2}$ for large $N$ [Fig.~\ref{fig:continuousscaling}(a)]. Thus the probability of observing Maxwell demon behavior in a continuous stochastic system is suppressed nearly exponentially as the number of degrees of freedom increases. The near-independent approximation~\eqref{eq:independentapprox} best fits the numerical sampling results, showing good agreement for all but $N=3$.

While the probability that a given random system is a Maxwell demon decreases rapidly with $N$, the expected value of the largest positive heat flow conditional on being a demon somewhat counterintuitively increases with $N$ [Fig.~\ref{fig:continuousscaling}(b)]. Moreover, the tail of the distribution for the largest heat flow becomes increasingly thick with larger $N$. Thus for small $N$ demons are likely, but rarely extract much heat from their environment. By contrast for large $N$ demons are exponentially unlikely, but when they do arise they can exhibit much stronger performance.

The analytic calculations detailed above rely on several simplifying assumptions about the dynamics, namely that the mobilities of each degree of freedom are all equal, that the diagonal terms of the force matrix $\A$ are all positive and equal, and that the off-diagonal terms of $\A$ are small relative to the diagonal terms ($\epsilon\ll1$). To test the robustness of our results we numerically sample random systems free from these simplifying assumptions (Appendix E provides details), and find that the probability $p(\mathrm{demon})$ is generally only smaller than suggested by our analytic calculations, and still decays at least exponentially with $N$.

\emph{Discrete dynamics}.---To validate our findings beyond linear Langevin dynamics, consider the case of discrete master equation dynamics. As a minimal null model, consider a system with $N$ binary degrees of freedom, where each $x_k \in \{0,1\}$. The state space is then an $N$-dimensional hypercube with $2^N$ states. To satisfy the multipartite assumption, transitions may occur only between states separated by a Hamming distance of 1, so that only one degree of freedom can transition at a time. To construct an ensemble, suppose that each entry of the transition rate matrix allowed to be non-zero is drawn independently from a log-normal distribution with $\ln w_{ij}\sim \mathcal{N}\left(0,\sigma_\mathrm{discrete}\right)$.

For a bipartite system ($N=2$) we solve for the probability of a Maxwell demon exactly, obtaining a probability of $1/2$ as in the case of continuous dynamics (Appendix C). For $N>2$, we take the linear response limit to make progress, assuming $\sigma_\mathrm{discrete}\ll 1$. As with the case of continuous dynamics, the probability for the $k$'th subsystem to have a positive heat flow can be mapped onto the alignment of vectors in high-dimensional space (detailed in Appendix D), such that
\begin{equation}
p_1 \approx  \frac{1}{2}I_{\frac{1}{2}}\left(\frac{2^{N-1} -1}{2},\frac{1}{2}\right),
\end{equation}
which scales for large $N$ as 
\begin{equation}
p_1\sim 2^{- 2^{N}-\frac{N}{2}}.
\end{equation} 
Unlike the continuous case this scaling is superexponential, because the dimensionality of the vectors that must align increases exponentially rather than linearly with $N$. From this marginal probability, just as in the continuous case we can determine both lower and upper bounds on the probability of the system as a whole exhibiting Maxwell demon behavior. The upper bound, corresponding to the most optimistic case for this probability, scales as
\begin{equation}
p_\mathrm{UB}(\mathrm{demon}|N) \sim N 2^{- 2^{N}-N/2},
\end{equation}
decaying as a double-exponential with $N$. We likewise formulate an independence approximation and a lower bound as before, both of which similarly decay double-exponentially with $N$.

\begin{figure}[t]
    \centering
    \includegraphics[width=\linewidth]{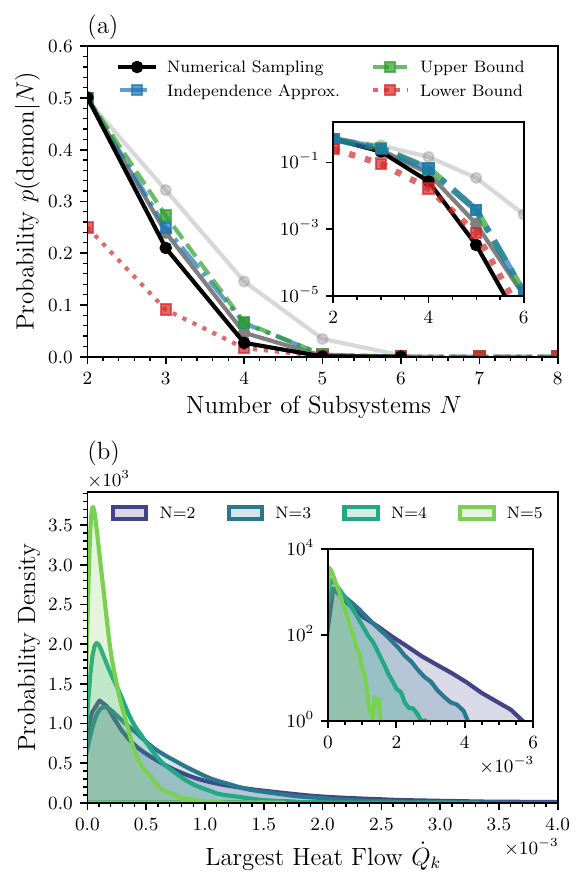}
    \caption{a) Probability that a random stochastic system obeying discrete master equation dynamics operates as a Maxwell demon. Theoretical lower (red) and upper (green) bounds, as well as the independence approximation (blue). Black (dark gray, light gray) points show numerical results from sampling random transition rate matrices with $\sigma_\mathrm{discrete}=0.1$ ($\sigma_\mathrm{discrete}=1,\, 10$) (uncertainties are smaller than the displayed points). Inset: logarithmic y-axis. b) Distribution of largest observed heat flows in random systems that are Maxwell demons for different numbers $N$ of degrees of freedom. Inset: logarithmic y-axis to illustrate tails of the distributions.
    }
    \label{fig:discretescaling}
\end{figure}

Numerically sampling from randomly generated transition matrices of size $2^N\times 2^N$ yields a probability of Maxwell demon behavior consistent with our calculations [Fig.~\ref{fig:discretescaling}(a)], confirming that for discrete state systems the probability of observing Maxwell demon behavior is superexponentially suppressed as the number of degrees of freedom increases. Indeed, by $N=6$ the probability of finding a demon has already decreased to $\mathcal{O}(10^{-6})$. While our analytic calculations require $\sigma_\mathrm{discrete}\ll1$, we numerically sample from ensembles free from this assumption, finding that superexponential scaling is robust (though the absolute probability $p(\mathrm{demon}|N)$ increases with $\sigma_\mathrm{discrete}$) . Thus we conclude that Maxwell demon behavior simply cannot arise by chance in large stochastic systems well-described by discrete-state master equation dynamics. Unlike in the continuous case, here the probability distribution for the largest heat flow given that a system is a demon shows both sharper peaks near $0$, as well as lighter tails, for larger $N$ [Fig.~\ref{fig:discretescaling}(b)]. Thus more degrees of freedom does not always enable stronger Maxwell demons. 

\emph{Discussion}.---In this letter we asked how likely it is that a random stochastic system will exhibit Maxwell demon behavior, with at least one subsystem extracting heat from its environment. Our results show that while this probability can be as high as $50\%$ for $N=2$, the most optimistic probability of finding a Maxwell demon decays exponentially with $N$ for continuous dynamics, and double-exponentially with $N$ for discrete dynamics. Thus we conclude that observing Maxwell demon behavior in a truly random large system is vanishingly small. These findings complement previous work showing that information flows scale sub-extensively with system size in CMOS implementations of Maxwell demons, such that the operation of demons becomes increasingly inefficient at macroscopic scales~\cite{freitas2022maxwell,freitas2023information}.

Intuitively, these results follow from the behavior of random vectors in high-dimensional space. As shown in our detailed derivations in the Appendix, for a multipartite system to operate as a Maxwell demon requires the alignment of two random isotropic vectors in $\mathbb{R}^\nu$, where $\nu$ scales linearly (continuous dynamics) or exponentially (discrete dynamics) with the number $N$ of subsystems. Such alignment becomes exponentially less likely as the dimensionality $\nu$ increases.

This finding has major implications. Most strikingly, our results show that we should be very surprised whenever we find evidence of Maxwell demon behavior in systems with many degrees of freedom. Since such behavior is highly unlikely to have arisen by pure chance, it stands to reason that some manner of selection must have taken place to produce the demon in question. Also surprisingly, we find in the case of linear Langevin dynamics that the magnitude of the largest positive heat flow in systems that do operate as demons increases with the number of degrees of freedom, with the tails of the distribution in particular becoming increasingly thick. Thus while demons become rarer with increasing $N$, the potential for particularly strong demons may also increase, though this result does not appear to extend to random systems with discrete dynamics.

Here we considered random models for both linear overdamped Langevin dynamics, which are generally in the near-equilibrium regime, and discrete-state master equation dynamics, which can be pushed arbitrarily far from equilibrium. In both cases, the probability of finding a demon decreases rapidly as the number of degrees of freedom increases, though the exact scaling relationship differs. Future work could consider modifications to the dynamics considered here, which we expect might change the quantitative scaling details, but is unlikely to change the qualitative result that demons become vanishingly unlikely as $N$ grows large. Another generalization of interest would be to systems in contact with reservoirs at different temperatures, a setting where Maxwell demon behavior has been shown to provide a performance advantage~\cite{leighton2024information}. 

Finally, we note that the definition we have used for a Maxwell demon, namely a multipartite system where at least one subsystem appears to locally violate the second law by taking in heat from the environment, is a relatively permissive definition compared to some in the literature with stricter requirements~\cite{sanchez2019nonequilibrium,ciliberto2020autonomous,saha2021maximizing,freitas2021characterizing}. As a result, the probabilities we have calculated here will more generally serve as upper bounds on the probabilities of finding demons under stricter definitions. Thus no matter how Maxwell's demon is defined, we should not expect to find it by chance in large random systems.

\emph{Acknowledgments}---We thank Jannik Ehrich (Deutsche Bahn) and David Sivak (SFU Physics) for helpful discussions, and Grégory Schehr for inspiring lectures on random matrix theory at the 2023 Beg Rohu Summer School. This work was supported by Mossman and NSERC Postdoctoral Fellowships.

\bibliography{main}

\appendix
\section*{Appendix}
\section{Langevin Dynamics with $N=2$}
We begin by studying the case of linear Langevin dynamics with $N=2$ degrees of freedom. In this case the force matrix $\A$ has only four terms, which we denote
\begin{equation}
\A = \begin{pmatrix} a & c\\ d & b\end{pmatrix}.
\end{equation}
The existence of a steady state requires both $\mathrm{Tr}(\A)>0$ and $\mathrm{det}(\A)>0$, implying $a+b>0$ and $ab>cd$. Solving the Lyapunov equation for $\cov$, we obtain exact expressions for the two heat flows. Without loss of generality, the first is $\dot{Q}_1  = c(d-c)/(a+b)$.
Since $a+b>0$, this heat flow will be positive if and only if $c(d-c)>0$, which in turn requires $c$ and $d$ to have the same sign, and satisfy $|d|>|c|$. Conversely, by symmetry the heat flow $\dot{Q}_2$ will be positive if and only if $c$ and $d$ have the same sign, and satisfy $|c|>|d|$. These events are thus disjoint, as expected, and overall the system operates as a demon if and only if $c$ and $d$ have the same sign. 

Suppose first that the diagonal terms are both positive and large relative to the off-diagonal terms, as assumed in the main text. Then $ab>cd$ is almost always true, and the existence of a steady state and existence of Maxwell demon behavior are independent. In that case if the off-diagonal terms $c$ and $d$ are drawn independently from a distribution symmetric about $0$, then it follows that the probability of obtaining a demon by randomizing the interaction matrix is exactly $1/2$.

As an alternate null model, we could instead place the diagonal and off-diagonal terms on equal footing by imposing that all of $a,b,c$, and $d$ are drawn independently from a distribution symmetric about $0$, and ask what is the probability of any such system that has a steady-state being a Maxwell demon. In this case, for the system to have a steady state and be a Maxwell demon requires all three of $a>0$, $b>0$, and $c$ and $d$ to have the same sign (all independent and each having probability $1/2$), and $ab>cd$, which given the first three also has probability $1/2$. Thus the total probability is $1/16$. The probability of a steady state existing, meanwhile, is $1/4$, so that the probability of a random system that has a steady state behaving as a Maxwell demon is $1/4$.

\section{Langevin Dynamics with $N\geq 2$}
For $N>2$, we restrict our attention to the case where $\A = \bm{I}+\epsilon \bm{M}$, with $\epsilon\ll1$, $M_{kk}=0$, and $M_{kj}\sim\mathcal{N}(0,1)$. In this case we solve the Lyapunov equation for the covariance by expanding in powers of $\epsilon$:
\begin{equation}
\cov = \cov_0 + \epsilon \cov_1  + \epsilon^2\cov_2 + \mathcal{O}(\epsilon^3).
\end{equation}
Calculating these terms, we obtain $\cov_0 = \bm{I}$,
\begin{subequations}
\begin{align}
\cov_1 & = -\frac{1}{2}\left(\bm{M} + \bm{M}^\top\right),\\
\cov_2 & = \frac{1}{4}\left[\bm{M}\left(\bm{M} + \bm{M}^\top\right) + \left(\bm{M} + \bm{M}^\top\right)\bm{M}^\top\right].
\end{align}
\end{subequations}
Inserting these into the heat flow $\dot{Q}_k$ [Eq.~\eqref{eq:continuousheatflow}], we obtain
\begin{equation}
\dot{Q}_k = \frac{\epsilon^2}{2} \sum_{j \neq k} \left( M_{kj}M_{jk} - M_{kj}^2 \right) + \mathcal{O}(\epsilon^3).
\end{equation}
A positive heat flow then requires the sum be greater than zero. To make progress, we note that this can be written in terms of two independent vectors in $\mathbb{R}^{N-1}$, namely $\bm{U} \equiv \{M_{kj}\}_{j\neq k}$ and $\bm{V}\equiv \{M_{jk}\}_{j\neq k}$, as the condition
\begin{equation}
\bm{U}\cdot\bm{V}> |\bm{U}|^2.
\end{equation}
Defining $V_\parallel$ to be the projection of $\bm{V}$ onto $\bm{U}$, we can simplify this condition to $V_\parallel >|\bm{U}|$. Note that $V_\parallel$ is a random variable following the standard normal distribution, while $|\bm{U}|$ is the square root of a $\chi^2$-distributed random variable with $\nu = N-1$ degrees of freedom. The distribution of their ratio,
\begin{equation}
t = \frac{V_\parallel}{|\bm{U}|/\sqrt{\nu}}
\end{equation}
is described by the Student's $t$ distribution $f_\nu(t)$. Thus the probability that $V_\parallel >|\bm{U}|$ is simply given by integrating the distribution function:
\begin{subequations}
\begin{align}
p(V_\parallel >|\bm{U}|) & = \int_{t=\sqrt{\nu}}^\infty f(t)\,\mathrm{d}t\\
& = \frac{1}{2} I_{\frac{1}{2}}\left(\frac{\nu}{2},\frac{1}{2}\right).
\end{align}
\end{subequations}
The resulting expression involving the incomplete regularized beta function is then the marginal probability $p_1$ that the $k$'th subsystem has a positive heat flow. 
For large $\nu$, this scales as $\nu^{-1/2}2^{-\nu/2}$.

\section{Discrete Dynamics with $N=2$}
For random systems obeying discrete-state master equation dynamics, we begin with the case $N=2$ where we can prove the result $p_1=1/4$ exactly. For $N=2$, there is exactly one cycle, and thus exactly one flux $J$ (equal for all four edges). The flux will have the same sign as the sum of forces on the edges from the two degrees of freedom,
\begin{equation}
\mathrm{sign}(J) = \mathrm{sign}(F_1 + F_2).
\end{equation}
For subsystem $1$ (without loss of generality) to have a positive heat flow, we must have $J F_1\leq 0$, which in turn requires both that $F_1$ and $F_2$ have different signs, and that $|F_2|>|F_1|$. Since these events are independent and each have probability $1/2$, the total probability $p_1$ is their product, equal to $1/4$. Thus for random discrete state dynamics with $N=2$ binary degrees of freedom, the probability of a Maxwell demon is exactly $1/2$.

\section{Discrete Dynamics with $N\geq2$}
Consider now $N\geq2$ binary degrees of freedom, where the state space is an $N$-dimensional hypercube with $2^N$ states. For a specific subsystem $k$, the transitions in which its state $x_k$ changes correspond to the ``edges'' of the $N$-dimensional hypercube in the $k$-th dimension. There are exactly $M = 2^{N-1}$ such edges connecting the subspace $x_k=0$ to $x_k=1$.

The total heat flow into subsystem $k$ is the sum of fluxes along all parallel transitions multiplied by the log ratios of transition rates~\cite{Ehrich2023_Energy}:
\begin{equation}
    \dot{Q}_k = -\sum_{\alpha=1}^{M} J_\alpha F_\alpha,
\end{equation}
where $\alpha$ indexes the $M$ edges corresponding to transitions of $x_k$. Here, $F_\alpha = \ln(w^+_\alpha / w^-_\alpha)$ is the thermodynamic force (log-ratio of forward/backward rates) on edge $\alpha$, and $J_\alpha = p_\alpha^+ w^+_\alpha - p_{\alpha}^- w^-_\alpha$ is the net probability current between the states (`$+$' and `$-$') on either end of the transition. 

Alternatively, the heat flow can also be written as a sum over all fundamental cycles $\gamma_k$ (see Ref.~\cite{schnakenberg1976network}) containing transitions in the $k$'th degree of freedom,
\begin{equation}
\dot{Q}_k = -\sum_{\gamma_k}J_{\gamma_k}F_{\gamma_k,k}.
\end{equation}
Here $J_{\gamma_k}$ is the flux for the cycle $\gamma_k$, and $F_{\gamma_k,k}$ is the sum of thermodynamic forces contributed from all edges in the cycle corresponding to transitions in $x_k$. Defining $F_{\gamma_k,\mathrm{ext}}$ as the sum of thermodynamic forces contributed from all cycle edges not corresponding to transitions in $x_k$, the cycle flux is generally given by
\begin{equation}
J_{\gamma_k} = \sigma_{\gamma_k} \left(e^{F_{\gamma_k,k} + F_{\gamma_k,\mathrm{ext}}} - 1\right).
\end{equation}
To make progress, we now make an approximation that the system is in the linear response regime, valid when $\sigma_\mathrm{discrete}\ll1$. In this case the exponential in the flux equation can be linearized, such that the heat flow is
\begin{equation}
\dot{Q}_k \approx -\sum_{\gamma_k}\sigma_{\gamma_k}\left(F_{\gamma_k,k}^2 + F_{\gamma_k,\mathrm{ext}}F_{\gamma_k,k}\right).
\end{equation}
The mobility $\sigma_{\gamma_k}$ depends on the full transition rate matrix, and thus for large $N$ will be independent of the forces $F_{\gamma_k,k}$ and $F_{\gamma_k,\mathrm{ext}}$. Thus we define the vectors $\bm{U} \equiv \sqrt{\sigma_{\gamma_k}}\{F_{\gamma_k,k}\}_{\gamma_k}$ and $\bm{V}\equiv -\sqrt{\sigma_{\gamma_k}}\{F_{\gamma_k,\mathrm{ext}}\}_{\gamma_k}$, and write the condition $\dot{Q}_k>0$ as
\begin{equation}
\bm{U}\cdot\bm{V}> |\bm{U}|^2.
\end{equation}
Assuming that the cycle forces are independent for any given pair of cycles, $\bm{U}$ and $\bm{V}$ are isotropic and independent random vectors with Gaussian-distributed entries in $\mathbb{R}^\nu$, where $\nu$ is the number of fundamental cycles containing transitions in the $k$'th degree of freedom. This now reduces to the same geometric problem considered in the case of continuous dynamics, so that the probability that the heat flow is positive is simply given by an incomplete regularized beta function,
\begin{equation}
p_1 = \frac{1}{2} I_{\frac{1}{2}}\left(\frac{\nu}{2},\frac{1}{2}\right),
\end{equation}
where $\nu$ is the number of degrees of freedom in the vectors. While it might seem t first glance that there should be $2^{N-1}$ degrees of freedom (corresponding to the number of edges), we must recall that there is a maximum number of independent cycles within the graph of transitions~\cite{schnakenberg1976network}, equal for an $N$-dimensional hypercube to
\begin{equation}
n_\mathrm{cycles}(N) = N2^{N-1} - 2^{N} + 1.
\end{equation}
Of these, only a subset involve transitions of the $k$'th degree of freedom. The number of irrelevant cycles is twice the number of cycles at each fixed value of $x_k$, equal to $2 n_\mathrm{cycles}(N-1)$. Thus the true number of degrees of freedom is
\begin{equation}
\nu = 2^{N-1} -1. 
\end{equation}

\section{Details of Numerical Sampling}
For continuous dynamics, we always work in dimensionless units where $T=1$ and $\langle\mu_{kk}\rangle=1$, and the entries of $\A$ are likewise dimensionless. For analytic tractability, we consider an ensemble where $\mu_{kk}=1$, and $\A = \bm{I} + \epsilon \bm{M}$ where $M={kk}=0$ and $M_{kj}\sim\mathcal{N}(0,1)$ for $k\neq j$. This gives the main numerical results in Fig.~\ref{fig:continuousscaling}. For robustness, we also sample from a more general ensemble where $\ln \mu_{kk} = \mathcal{N}(0,1)$, and $\A_{ij}\sim \mathcal{N}(0,1)$ for all $i,j$. This gives the gray points and curve in Fig.~\ref{fig:continuousscaling}(a). In both cases we reject any samples that do not produce steady states, requiring all eigenvalues of $\A$ be positive. 

For discrete dynamics, we sample the allowed transition rates $w_{ij}$, where states $i$ and $j$ have Hamming distance $1$, from a log-normal distribution such that $\ln w_{ij}\sim\mathcal{N}(0,\sigma_\mathrm{discrete})$. Our analytic results are compared with the $\sigma_\mathrm{discrete}=0.1$ ensemble, while for robustness we also numerically sample $\sigma_\mathrm{discrete}=1,10$.

\end{document}